\documentclass[12 pt, draftcls]{IEEEtran}\onecolumn

\usepackage[dvips]{color}
\usepackage{epsf}
\usepackage{times}
\usepackage{epsfig}
\usepackage{graphicx}
\usepackage{amsmath}
\usepackage{amssymb}
\usepackage{amsxtra}
\usepackage{here}
\usepackage{subfigure}
\usepackage{rawfonts}
\usepackage{times}
\usepackage{url}
\usepackage{cite}
\usepackage{algorithm}

\newtheorem{defn}{\bf Definition}

\IEEEoverridecommandlockouts

\begin{document}

\title{Matching-based Spectrum Allocation in Cognitive Radio Networks}

\author{\IEEEauthorblockN{Raghed El-Bardan\IEEEauthorrefmark{1}, Walid Saad\IEEEauthorrefmark{2}, Swastik Brahma\IEEEauthorrefmark{1}, and Pramod K. Varshney\IEEEauthorrefmark{1}}\\
\IEEEauthorblockA{\IEEEauthorrefmark{1}Dept. of Electrical Engineering and Computer Science, Syracuse University, Syracuse, NY, $\{$raelbard, skbrahma, varshney$\}$@syr.edu} \\  \IEEEauthorrefmark{2}Wireless@VT, Bradley Dept. of Electrical and Computer Engineering, Virginia Tech, Blacksburg, VA, walids@vt.edu
\thanks{This research was supported by CASE: A NYSTAR Center for Advanced Technology and the U.S. National Science Foundation under Grants CNS-1460316, CNS-1460333, CNS-1443913, CNS-1443914, and AST-1443913.}}

\maketitle
\thispagestyle{empty}
\pagestyle{empty}

\begin{abstract}
In this paper, a novel spectrum association approach for cognitive radio networks (CRNs) is proposed. 
Based on a measure of both inference and confidence as well as on a measure of quality-of-service, the association between secondary users (SUs) in the network and frequency bands licensed to primary users (PUs) is investigated. 
The problem is formulated as a \emph{matching game} between SUs and PUs. In this game, SUs employ a soft-decision Bayesian framework to detect PUs' signals and, eventually, rank them based on the logarithm of the \textit{a posteriori} ratio. A performance measure that captures both the ranking metric and rate is further computed by the SUs. Using this performance measure, a PU evaluates its own utility function that it uses to build its own association preferences. A distributed algorithm that allows both SUs and PUs to interact and self-organize into a \emph{stable match} is proposed. Simulation results show that the proposed algorithm can improve the sum of SUs' rates by up to $20 \%$ and $60 \%$ relative to the deferred acceptance algorithm and random channel allocation approach, respectively. The results also show an improved convergence time.
\end{abstract}


\IEEEpeerreviewmaketitle

\section{Introduction}
\label{sec:Introduction}

The proliferation of new wireless technologies has led to an increasing demand for the scarce radio spectrum resources, thus motivating operators and governmental agencies to rethink the way in which existing fixed spectrum allocation policies are defined \cite{FCCtech}. 
In this context, cognitive radios, based on dynamic spectrum access (DSA) techniques, have emerged as a promising communication paradigm for deploying flexible and adaptive spectrum management mechanisms~\cite{CR}. 
Cognitive radio networks (CRNs) involve two classes of users: primary users (PUs), who are the incumbent licensees of the spectrum, and unlicensed secondary users (SUs), who either sense the spectrum and transmit over vacant bands or operate on licensed spectrum bands as long as their transmissions do not cause harmful interference to the activities of PUs. In order to reap the benefits of CRNs, it is imperative to design smart and agile spectrum allocation mechanisms that can achieve better management of spectral resources. 

For spectrum allocation in a CRN, two fundamental architectures have recently been studied~\cite{surveyDSA-1,surveyDSA-2,Hussein2009,El-Bardan2013}. One is centralized where a central entity, such as a central controller or spectrum broker, is in charge of allocating the spectrum or part of it to different SUs~\cite{CenScheme-4}. Based on this scheme, SUs forward their spectrum sensing measurements to a central entity which constructs a spectrum allocation map. Naturally, a centralized approach can lead to high communication overhead and is not scalable. 
Alternatively, a distributed architecture can be adopted, for example, when the construction of a centralized infrastructure is not possible and/or for quick adaptation to network dynamics. 
Therefore, we focus our attention on distributed architectures in this paper.
 
Distributed spectrum allocation, in the context of wireless networks, has received considerable attention recently ~\cite{DisResManCRNDST,DistSpecAllocBarg,MulChConCRS,rami2014}. 
The authors in \cite{DisResManCRNDST} developed a distributed algorithm for multi-user channel allocation in CRNs that is based on the learning of the behavior of PUs through a multi-agent learning concept. However, in \cite{DisResManCRNDST}, for the network nodes to learn and take decisions, they must have precise and timely information such as channel information or interference patterns that might be hard to gather in a distributed setting. 
In \cite{DistSpecAllocBarg}, the authors introduced an adaptive and distributed local bargaining approach where mobile users self-organize into bargaining groups and adapt their previous spectrum assignment to approximate a new optimal assignment. Nevertheless, their approach is highly centralized. 
The work in \cite{MulChConCRS} studies the problem of channel allocation in wireless networks using a matching theory-based mechanism that is solved via a variant of the so-called Gale-Shapley deferred acceptance algorithm \cite{Gale-shap}. They analyze the performance of the proposed solution from the user's perspective and provide tight lower and upper bounds on both the stable allocation and the optimal centralized allocation performance. In \cite{rami2014}, the authors use matching theory to find a stable multi-channel allocation for each SU assuming the existence of a central coordinator responding on behalf of the PUs. They investigate cooperative channel assignments that require direct communication between SUs and prove the existence of an equilibrium solution. However, in \cite{rami2014}, the scheme is not fully distributed and it incurs additional communication overhead. 

The main contribution of this paper is to design a novel and distributed spectrum allocation approach in a CRN when SUs have non-identical spectrum uncertainty values. We consider a CRN in which each SU perceives the availability of the spectrum differently. 
Inspired by the stable marriage problem \cite{Gale-shap}, we formulate the problem as a \emph{matching game} in which the SUs and PUs are the players that need to evaluate one another in order to find suitable associations (an association is an allocation of a PU-owned frequency band to a unique SU). From the SU's perspective, the ranking is done based on the inference and confidence measures regarding the presence or absence of PUs over the licensed spectrum. Prior to proposing to the PUs, each SU computes an overall performance measure (utility) that reflects its valuation of the bands by capturing the effect of rate and its ranking metric. From the PU's perspective, the ordering is dependent on the SUs' proposals as we, in contrast to \cite{Gale-shap}, assume that PUs are not able to rank SUs on their own. In other words, using the values of the overall performance measure proposed by the SUs, a PU evaluates its utility function that it uses to determine its association preferences. 
The key advantage of the proposed approach lies in the fact that the SU-PU association is achieved through distributed decisions at each SU and PU. To solve the proposed matching game, a novel distributed algorithm is introduced. This algorithm enables the SUs and PUs to self-organize into a \emph{stable match}. Simulation results show that the proposed algorithm can improve the sum of SUs' rates by up to $20\%$ and $60\%$ relative to the deferred acceptance algorithm and random channel allocation approach, respectively. The results also show an improved convergence time. 

Compared to existing works such as \cite{DistSpecAllocBarg} and \cite{rami2014}, the proposed approach is fully distributed as it requires no cluster-head to drive the assignment, no central entity to perform the bargaining, no coordinator to control the matching on behalf of the PUs, and no information exchange among SUs. In contrast to \cite{DisResManCRNDST}, the proposed algorithm neither requires the learning of users in the network nor timely and precise information about them. Different from \cite{MulChConCRS} and \cite{rami2014}, we consider that SUs have different information regarding the activity of PUs over their licensed bands. Compared to \cite{MulChConCRS}, our proposed scheme enables the frequency band licensees in the network to be active players in the association process and, thus, make better informed spectrum association decisions that maximize their own payoffs.   
In addition, the performance analysis we provide is not based on bounds.  
In contrast to \cite{rami2014} where the authors assume the existence of a central coordinator that responds to offers from all the SUs on behalf of the PUs, we investigate a one-to-one stable match for which we propose a completely distributed resource allocation algorithm. 

The rest of this paper is organized as follows. Section \ref{sec:SystemModel} introduces the system model while Section \ref{sec:ProbForm} presents the game formulation and properties. In Section \ref{sec:SpecAlg}, we present the solution to the game and propose a distributed algorithm based on matching theory. Simulation results are provided in Section \ref{sec:NumRes}. Finally, Section \ref{sec:Conclusion} summarizes our work.  
  
\section{System Model}
\label{sec:SystemModel}

\begin{figure}[t]
\centering
\includegraphics[width=0.475\textwidth,height=0.215\textwidth]{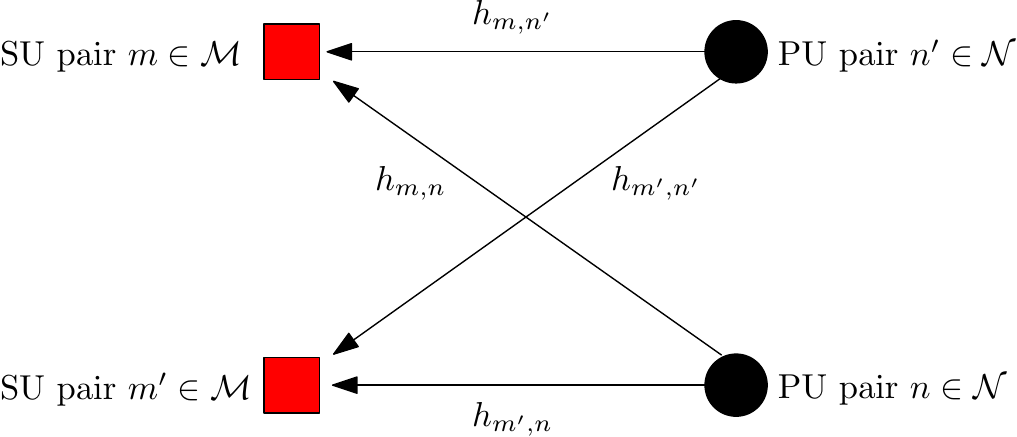}
\caption{\small An illustration of the system model. An arrow portrays the sensing mechanism of PUs activities performed at the transmitter's side of each SU pair.}
\label{fig:net-model}
\end{figure}
Consider a single-hop CRN consisting of a set $\mathcal{M} = \{1,\cdots,M\}$ of $M$ SU transmitter-receiver pairs and a set $\mathcal{N} = \{1,\cdots,N\}$ of $N$ PU transmitter-receiver pairs, with $M > N$\footnote{We consider this realistic scenario for illustration purposes only. However, our work can be easily extended to $M \le N$.}. For brevity, we use the term SUs and PUs to denote, respectively, the SU and PU transmitter-receiver pairs. 
PUs are licensed users whose transmissions are slowly varying over time. Each PU operates over $1$ out of $N$ orthogonal licensed frequency bands through which transmissions are assumed to be collision-free. At any point in time, a given PU may be active or inactive. 
However, from the perspective of any SU $m \in \mathcal{M}$, every PU $n \in \mathcal{N}$ is considered to be active with a probability $\pi_{m,n}$. For a given PU $n \in \mathcal{N}$, two distinct SUs $m$, $m' \in \mathcal{M}$, $m \neq m'$ may have a different value of the probability that $n$ is active, i.e., we may have $\pi_{m,n} \neq \pi_{m',n}$, depending on various factors such as the distance to the PU and the wireless channel state. Let $\boldsymbol{\pi}_m = [\pi_{m,1} \cdots \pi_{m,N}]$ for any $m \in \mathcal{M}$. Furthermore, the wireless channel model through which signals are transmitted is assumed to be frequency selective. We take into account a broad class of channel models which consist of a known distance-independent frequency selective component, and a deterministic distance-dependent path loss component with path loss exponent $\gamma$ \cite{Weber2007}. More specifically, let
\begin{eqnarray}
h_{m,n} = \sqrt{\frac{\beta_{n}}{1 + k d_{m,n}^{\gamma}}} 
\end{eqnarray} be the channel gain over a distance $d_{m,n}$, between the transmitters of SU $m$ and PU $n$, with a deterministic frequency selective coefficient $\beta_{n}$ and a constant $k$. Note that, for any two PU-owned frequency bands $n$ and $n'$, $\beta_{n} \neq \beta_{n'}$. 
Note that this path loss model is consistent with the one employed in \cite{Varshney2004} and it is valid even if $d_{m,n}$ is close to or equal to $0$. 

Furthermore, we let $\boldsymbol{\eta}_m = [\eta_{m,1} \cdots \eta_{m,N}]$ for any $m \in \mathcal{M}$ denote the rates over the bands in $\mathcal{N}$ where $\eta_{m,n}$, for $n \in \mathcal{N}$, is given by \cite{goldsmith_wc}:
\begin{equation}\vspace{-.25cm}
\eta_{m,n} = \log_2\left(1 + \frac{P_{T_m}g_{m,n}^2}{\sigma^2}\right), \quad \forall m, n, \label{achievable-rate} 
\end{equation} where $P_{T_m}$ denotes SU $m$'s transmit power, $g_{m,n} = \sqrt{\frac{\beta^{'}_{n}}{1 + k d_{m,m}^{\gamma}}}$ is the channel gain of PU-owned band $n$ over a distance $d_{m,m}$ between SU transmitter-receiver pair $m$ with a known frequency selective coefficient $\beta^{'}_{n}$, and $\sigma^2$ is the additive white Gaussian noise variance, assumed the same for all SUs over all bands.

In the model studied here, SUs can only communicate with PUs. In other words, there does not exist any negotiation or information exchange among SUs. We further assume that an SU is capable of transmitting over a single PU-owned frequency band at a time, but can sense the activities of all PUs in $\mathcal{N}$. In this paper, the system is assumed to be slotted where each SU $m \in \mathcal{M}$ makes a single observation every $T$ time slots and decides on the presence or  absence of a PU $n \in \mathcal{N}$ on its licensed channel based on this observation.

Local sensing for primary signal detection, done once every $T$ time slots, can be formulated as a binary hypothesis testing problem as follows \cite{crahns}:
\begin{eqnarray}
\mbox{under } H_0: x_{m,n} &=& w, \nonumber \\
\mbox{under } H_1: x_{m,n} &=& h_{m,n}s_{n} + w,\nonumber
\end{eqnarray}
where, at SU $m \in \mathcal{M}$ over PU-owned frequency band $n \in \mathcal{N}$, $x_{m,n}$ denotes the received signal, $h_{m,n}$ is the $n$-th channel gain, $s_{n}$ denotes the known signal of PU $n$, and $w$ is the zero-mean additive white Gaussian noise. $H_0$ and $H_1$ denote the absence and the presence, respectively, of the PU signal in the frequency band. 

In our model, we consider the above signal detection problem under a soft-decision Bayesian framework. We compute the logarithm of the \textit{a posteriori} ratio that captures both inference and confidence measures. In other words, the sign of the logarithm of the \textit{a posteriori} ratio yields the decision on $H_1$ or $H_0$ while its magnitude determines the confidence regarding a decision on either one of the two hypotheses. More formally, we present this statistic of the aforementioned binary hypothesis testing problem as follows:
\begin{eqnarray}
\delta_{m,n} &=& \log \left(\frac{P\left(H_1|x_{m,n}\right)}{P\left(H_0|x_{m,n}\right)}\right) \nonumber \\ &=& \log \left(\frac{\pi_{m,n}p\left(x_{m,n}|H_1\right)}{\left(1 - \pi_{m,n}\right)p\left(x_{m,n}|H_0\right)}\right) \mathop{\stackrel{H_1}{\gtrless}}_{H_0} 0, \label{loglike}
\end{eqnarray} where $P\left(H_1|x_{m,n}\right)$ and $P\left(H_0|x_{m,n}\right)$ denote the \textit{a posteriori} probabilities based on the  SUs' observations. $p\left(x_{m,n}|H_1\right)$ and $p\left(x_{m,n}|H_0\right)$ represent the likelihood functions of $x_{m,n}$ as perceived by the SUs under $H_1$ and $H_0$ respectively, and, are given by
\begin{eqnarray}
p\left(x_{m,n}|H_1\right) &=& 
 \frac{1}{\sigma \sqrt{2 \pi}}e^{-\frac{\left(x_{m,n} - h_{m,n} s_{n}\right)^2}{2\sigma^2}}, \label{eq:pos_H1} \\
p\left(x_{m,n}|H_0\right) 
&=&\frac{1}{\sigma \sqrt{2 \pi}}e^{-\frac{\left(x_{m,n}\right)^2}{2\sigma^2}}.\label{eq:pos_H0}
\end{eqnarray} 
It has been shown that likelihood ratio test based detection schemes are optimal, i.e., the probability of error is minimized in a Bayesian framework\cite{Varshney1996}. Therefore, using the logarithm of the \textit{a posteriori} ratio in our model captures the confidence regarding the presence or absence of PU activity over a particular band. In fact, this measure constitutes the SU's ranking metric which, in turn, is used to order the PU-owned frequency bands in $\mathcal{N}$. For a given SU $m \in \mathcal{M}$ and two distinct PU-owned frequency bands $n$, $n' \in \mathcal{N}$, $n \neq n'$ and $\delta_{m,n} < \delta_{m,n'}$, it prefers $n$ over $n'$. As a result, the SU would use PU-owned frequency band $n$ for its transmissions. It is, however, possible that the same band is the preferred choice of another SU. Hence, competitions exist among SUs for PU-owned bands and the problem of finding a stable allocation of each PU-owned band in $\mathcal{N}$ to a unique SU in $\mathcal{M}$ arises. 

In our model, we assume that SUs as well as PUs participate in the association process. However, PUs need an incentive to grant SUs access to their licensed spectrum. In this regard, for any SU $m \in \mathcal{M}$ that is looking to transmit over any PU-owned frequency band $n \in \mathcal{N}$, we define $v_{m,n}$, a function of both the ranking metric and rate, to serve as a measure of SU $m$'s utility: 
\begin{eqnarray}
\label{payoff-SUU}
v_{m,n} = f\left(\delta_{m,n}, \eta_{m,n}\right).
\end{eqnarray}
Using the utility of an SU, a PU evaluates its utility function. Accordingly, we define $u_{n,m}$ 
as PU $n$'s utility when associated with SU $m$. It is assumed, in this framework, that $u_{n,m}$ 
is monotonically increasing with $v_{m,n}$ for PU $n$ that is inactive and whose licensed band is available. One possible way of interpreting the aforementioned assumption is to consider $u_{n,m}$
as a reward or credit that the SU proposes to give to the inactive PU based on the former's valuation of the latter's frequency band. However, if PU $n$ is active, we assume $u_{n,m} = 0$ $\forall m$ as it values its own transmission more than all SUs' proposals and, thus, it does not associate with any SU. This utility function serves as the incentive for a PU to participate in the association process and determines its preferences. For instance, a PU that is inactive prefers to allocate its licensed band to an SU with high utility, e.g., an SU which is more likely to detect it as inactive and attains a high rate over the licensed band so that the PU gets a higher reward. An active PU, on the other hand, prefers to keep the licensed band for its own transmissions. 
Here, we note that the PUs in the network do not have information about the SUs' inference and confidence levels as well as their rates. Hence, the utilities of SUs, $v_{m,n}$ for all $m$ and $n$, is unknown to the PUs initially. In other words, inactive PUs are unable to evaluate their utilities in the beginning and, therefore, cannot make association decisions that maximize their payoffs. 
In this case, an inactive PU $n$ can choose an SU $m$ to associate with only when it acquires information on the latter's utility, $v_{m,n}$. 
This is the main reason we require SUs to forward their proposals to PUs in our system according to their preference lists. 

This said, our goal is to obtain a spectrum allocation strategy according to which each SU is associated with its most preferred PU-owned frequency band and vice versa. To do this, we formulate the problem as a \emph{matching game} between PUs and SUs. We first define the game in Section \ref{sec:ProbForm} and, then, present the solution concept and propose a distributed resource allocation algorithm in Section \ref{sec:SpecAlg}. 

\section{Spectrum Allocation as a Matching Game}
\label{sec:ProbForm}

Using classical optimization techniques to solve the frequency allocation problem can yield significant overhead. As a matter of fact, it is NP-hard in general \cite{nphard-prob}. Therefore, the need for self-organizing solutions in CRNs along with the complexity of centralized optimization methods necessitate a distributed framework in which SUs and PUs autonomously determine, based on their individual objectives, the best SU-PU associations. One suitable mechanism for developing such an autonomous SU-PU associations approach is given by \emph{matching theory} \cite{matching-book}.

In this section, we are interested in finding SU-PU associations based on which any available PU-owned frequency band is granted to the best SU in $\mathcal{M}$ through a matching $\xi$: $\mathcal{N} \rightarrow \mathcal{M}$. 

\subsection{Matching Concepts}
\label{sec:MatchConcept}
  
A matching is defined as an assignment of SUs in $\mathcal{M}$ to PUs in $\mathcal{N}$ where each of the available bands is allocated exclusively to one SU. In this case, $M-N$ or more SUs will not be matched. We denote by $\mathcal{K}$ the set of available licensed bands where $\mathcal{K} \subseteq \mathcal{N}$ and $\mathcal{S}$ the set of matched SUs where $\mathcal{S} \subset \mathcal{M}$. 
More formally, a matching game can be defined as in \cite{Gale-shap}:
\begin{defn}
\label{defn1}
A \emph{matching game} is defined by two sets of players ($\mathcal{M}$, $\mathcal{N}$) and two preference relations $\succ_m$ and $\succ_n$. $\succ_m$ allows players of set $\mathcal{M}$ to evaluate (rank) players of set $\mathcal{N}$, while $\succ_n$ allows players of set $\mathcal{N}$ to rank players of set $\mathcal{M}$. 
\end{defn} 
The outcome of a matching game is a one-to-one association function $\xi$ that bilaterally assigns to each player $m \in \mathcal{S}$, a player $n = \xi(m)$, $n \in \mathcal{K}$. Similarly, we have $m = \xi(n)$. To complete the definition of the game, we must introduce a preference relation $\succ$ which is defined as a complete, reflexive, and transitive binary relation between the players in $\mathcal{M}$ and $\mathcal{N}$. Thus, for any SU $m \in \mathcal{M}$, a preference relation $\succ_m$ is defined over the set of PUs $\mathcal{N}$ such that, for any two PU-owned frequency bands $n$, $n' \in \mathcal{N}$, $n \neq n'$:
\begin{equation}
\label{pref-m}
n \succ_m n' \Leftrightarrow \delta_{m,n} < \delta_{m,n'},
\end{equation} where $\delta_{m,n}$ and $\delta_{m,n'}$ denote the payoff of SU $m$ corresponding to $n$ and $n'$ respectively.      
Similarly, for any available PU-owned frequency band $n$, a preference relation $\succ_n$ over the set of SUs $\mathcal{M}$ is defined as follows, for any two SUs $m$, $m' \in \mathcal{M}$, $m \neq m'$:
\begin{equation}
\label{pref-n}
m \succ_n m' \Leftrightarrow u_{n,m}\left(v_{m,n}\right) > u_{n,m'}\left(v_{m',n}\right),
\end{equation} where $u_{n,m}$ and $u_{n,m'}$ denote PU $n$'s utility when it is inactive and grants its frequency band to SU $m$ and $m'$ respectively. 

According to the preference relation defined in (\ref{pref-m}), we will next show how SUs build their preferences and, accordingly, evaluate their proposals. Then, in Section \ref{sec:PU-pref}, we present how PUs play an active role in the association process. 

\subsection{Preferences of the SUs}
\label{sec:SU-pref}  

So far, the logarithm of the \textit{a posteriori} ratio for each of the SUs in $\mathcal{M}$ has been computed according to (\ref{loglike}), (\ref{eq:pos_H1}), and (\ref{eq:pos_H0}). Based on the results obtained, an SU $m$ ranks the PU-owned frequency bands in an increasing order with the largest element being the least preferred. 
Accordingly, an SU $m$ obtains a preference list $\boldsymbol{\Delta}_m$, which is sorted based on (\ref{pref-m}). 

Given $\boldsymbol{\Delta}_m$ and $\boldsymbol{\eta}_m$, an SU $m \in \mathcal{M}$ is able to better evaluate the benefit from using each of the PU-owned frequency bands. Intuitively, in addition to $\boldsymbol{\eta}_m$, an SU values a frequency band less if it predicts the band to be occupied by a PU, i.e., $\delta_{m,n} > 0$. On the other hand, an SU values a frequency band more if it predicts the band to have no PU activity on it. In other words, when $\delta_{m,n} < 0$, $v_{m,n}$ in (\ref{payoff-SUU}) is higher than when $\delta_{m,n} > 0$. Hence, $v_{m,n}$ should be defined in a way such that the aforementioned properties are satisfied. We, in this paper, express it as a weighted sum of the ranking metric and rate. More precisely, an SU $m$'s utility that it proposes to PU $n$ is given by:     
\begin{eqnarray}
v_{m,n} \left(\delta_{m,n}, \eta_{m,n}\right) &=& -\alpha_m \delta_{m,n}  + \left(1 - \alpha_m\right) \eta_{m,n}, \label{eq:propsal-vec} 
\end{eqnarray} where $\alpha_m$ is a positive weight chosen by SU $m$ such that $0 \le \alpha_m \le 1$. 
Next, we present the perspective of PUs on the matching game.      
 
\subsection{Preferences of the PUs}
\label{sec:PU-pref}  

Initially, an inactive PU $n \in \mathcal{K}$ has no known preferences as to which SU it should grant its licensed band to. It is, however, assumed that its corresponding utility function is monotonically increasing with the utility of the SU it is associated with. When the matching game starts, only inactive PUs start to build their preferences based on the proposals they receive from SUs in the form of their utility functions. In the case where the utilities of SUs are non-negative, i.e., $v_{m,n} > 0$ $\forall m,n$, every inactive PU will have a preferred SU while every active PU will keep the licensed band to itself. However, it is possible for an inactive PU $n$ not to be matched. This may happen in the following two scenarios: i) $v_{m,n} < 0$ $\forall m$, or ii) a restriction on the number of SUs' proposals to be made exists\footnote{This scenario is beyond the scope of this paper.}.

Inactive PUs are interested in getting associated with SUs presenting the highest value of overall performance measure. In this manner, an inactive PU maximizes its utility. In fact, it accepts or rejects SUs' proposals based on the preference relation defined in (\ref{pref-n}) and the fact that its utility is a monotonically increasing function with that of the SU. An active PU, on the other hand, simply rejects all the SUs' proposals. 

So far, we have defined the matching game and presented how SUs build their preferences and proposals. We have also shown how PUs adopt their preferences. 
Next, we define stability, the solution approach for this matching game, and, propose a novel distributed algorithm that finds a stable spectrum allocation. 

\section{Game Solution and Proposed Algorithm}
\label{sec:SpecAlg}

Having defined the preference relations of the SUs in (\ref{pref-m}) and the inactive PUs in (\ref{pref-n}) as well as articulated the  modus operandi of active PUs, we, next, characterize the matching between SUs and PUs. One suitable solution for the proposed matching game is the so-called stable match defined as in \cite{Gale-shap}:
\begin{defn}
A matching $\xi$ with any SU-PU association $\left(m,n\right) \in \xi$, $m \in \mathcal{S}$ and $n \in \mathcal{K}$, is said to be \emph{stable} if there does not exist any SU $m'$ or PU $n'$, for which SU $m$ prefers PU $n'$ over PU $n$, or any PU $n$ which prefers SU $m'$ over SU $m$. 
\end{defn}
\begin{algorithm} [t] 
\small                       
\caption{SU-PU Associations Algorithm}          
\label{alg-alloc}                           
\textbf{Data}: Initially, no frequency band is assigned to any SU.\\
\textbf{Result}: A stable spectrum allocation (matching) $\xi$.\\
\textbf{Phase I - Detection of PUs Signals}
\begin{itemize}
\item Each SU $m$ knows $\boldsymbol{\pi}_m$.
\item Following a soft-decision Bayesian detection framework, each SU $m$ computes the logarithm of the \textit{a posteriori} ratio over frequency band, $n$ $\delta_{m,n}$, $\forall m, n$.  
\end{itemize}
\textbf{Phase II - Building SUs Preferences and Proposals}
\begin{itemize}
\item An SU $m$ constructs its preference vector $\boldsymbol{\Delta}_m$ based on $\delta_{m,n}$, $\forall n$. $\boldsymbol{\Delta}_m$ is sorted based on $\succ_m$.
\item Each SU $m$ constructs its proposals, as shown in (\ref{eq:propsal-vec}) for all $n$, and keeps it only if $v_{m,n} > 0$. 
\end{itemize}
\textbf{Phase III - Matching Evaluation}

Initially, all PUs have no preferred SU.\\
\textbf{repeat}\\
\hspace*{.25cm} \textbf{if} $n$ $\succ_m$ $n'$, then \\
\hspace*{.50cm} An SU $m$ sends a proposal to PU $n$.\\
\hspace*{.50cm} PU $n$ receives its proposal and, if active, rejects it.\\
\hspace*{.50cm} Otherwise, PU $n$ accepts it if it hasn't been proposed to before.\\
\hspace*{.50cm} When PU $n \in \mathcal{K}$ and it has been proposed to before,\\ 
\hspace*{.75cm} \textbf{if} $m$ $\succ_n$ $m'$ ($v_{m,n} > v_{m',n}$ and $u_{n,m}(v_{m,n}) > u_{n,m'}(v_{m',n})$)\\ \hspace*{1cm} PU $n$ rejects $m'$ proposal and adopts $m$'s. \\
\hspace*{.75cm} \textbf{else}\\ \hspace*{1cm} PU $n$ keeps the proposal he has and rejects the new one.\\ \hspace*{.75cm} \textbf{end}\\
\hspace*{.25cm} \textbf{end}\\ 
\textbf{until} All positive proposals have been made and a stable match is reached.\\
\textbf{Phase IV - Spectrum Allocation} 
\end{algorithm} 
In order to reach a stable match, we propose Algorithm \ref{alg-alloc}, a variant of the deferred acceptance algorithm, that consists of four main phases: detection of PUs' signals, building SUs' preferences and constructing proposals, matching evaluation, and finally SU-PU associations. 

Initially, each SU $m$ computes the logarithm of the \textit{a posteriori} ratio over PU-owned frequency band $n$ ($\delta_{m,n}$, $\forall m, n$). In the second phase, each SU $m$ obtains its preference vector $\boldsymbol{\Delta}_m$ by ranking $\delta_{m,n}$ based on (\ref{pref-m}) $\forall n$. Then, using $\boldsymbol{\eta}_m$ and $\boldsymbol{\Delta}_m$, the SU evaluates $v_{m,n}$ defined in (\ref{eq:propsal-vec}) for all PU-owned frequency bands. Different from the well-known deferred acceptance algorithm, an SU discards all the proposals with a negative performance measure, $v_{m,n}$, from its preference list. That is, an SU in the network does not propose to a PU-owned frequency band for which $v_{m,n} < 0$. This is because we, in this paper, consider that an SU's utility over a PU-owned band has to be non-negative in order for the SU to participate in the association process and, accordingly, propose to the PU. If an SU $m$ is not currently allocated the most preferred frequency band $n$, it sends $n$ a matching proposal. Upon receiving a proposal, PU $n$, if active, rejects the proposal. Otherwise, PU $n$ updates its utility and accepts the request of the SU either when it has not been proposed to yet or when the utility is expected to be greater than what it gets from another SU's request. If the proposal is rejected, SU $m$ proposes to the next PU-owned frequency band in its preference list. Once the algorithm terminates, a stable match, that consists of $N$ SU-PU associations, is reached as long as all PUs are inactive and there does not exist a scenario where, for an inactive PU $n$, we have $v_{m,n} < 0$ $\forall m$. Otherwise, the number of associations in the stable match is less than $N$. In this case, there are $M-N$ or more SUs who are not associated with PUs. The existence of such a stable frequency allocation strategy is conditioned on the SUs forwarding the necessary proposals to PUs. Once these proposals have been made, each active PU keeps its licensed band while each inactive PU remains associated with the most preferred SU and Phase III terminates. 

Our framework investigates one-to-one SU-PU associations. In this case, the existence of a stable match is guaranteed by \cite{Gale-shap} where Gale and Shapley proved its existence for any standard game. The convergence of Algorithm \ref{alg-alloc} follows from that of Phase III which is based on \cite{Gale-shap}. 

\section{Numerical Results}
\label{sec:NumRes}

\begin{figure}[t]
\centering
\includegraphics[width=0.65\textwidth,height=0.45\textwidth]{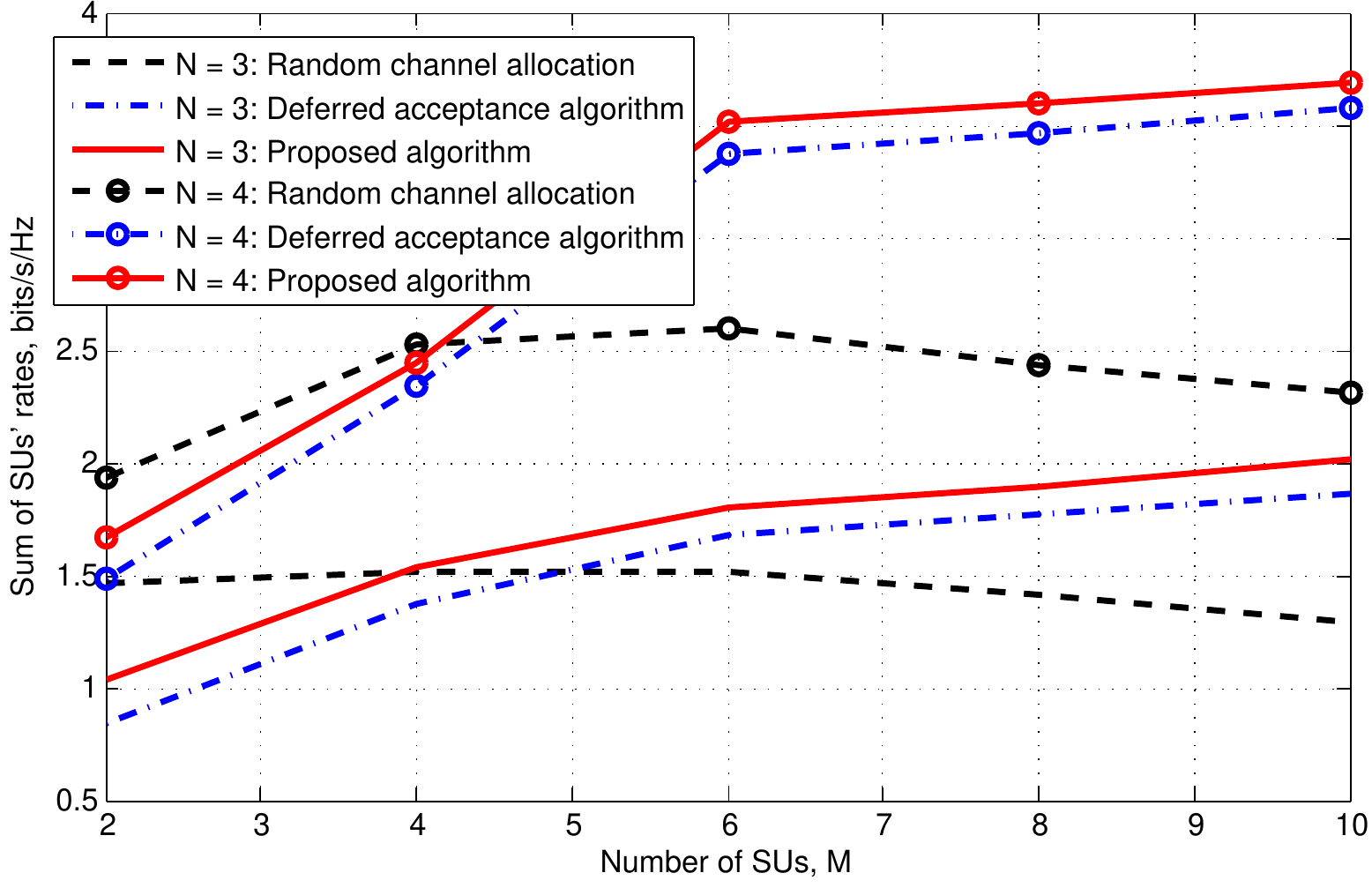}
\caption{\small The sum of SUs' rates with respect to the number of SUs in the network, $M$, for $N$ = $\{3, 4\}$.}
\label{fig:throughput-SUs}
\end{figure}
\begin{figure}[t]
\centering
\includegraphics[width=0.65\textwidth,height=0.45\textwidth]{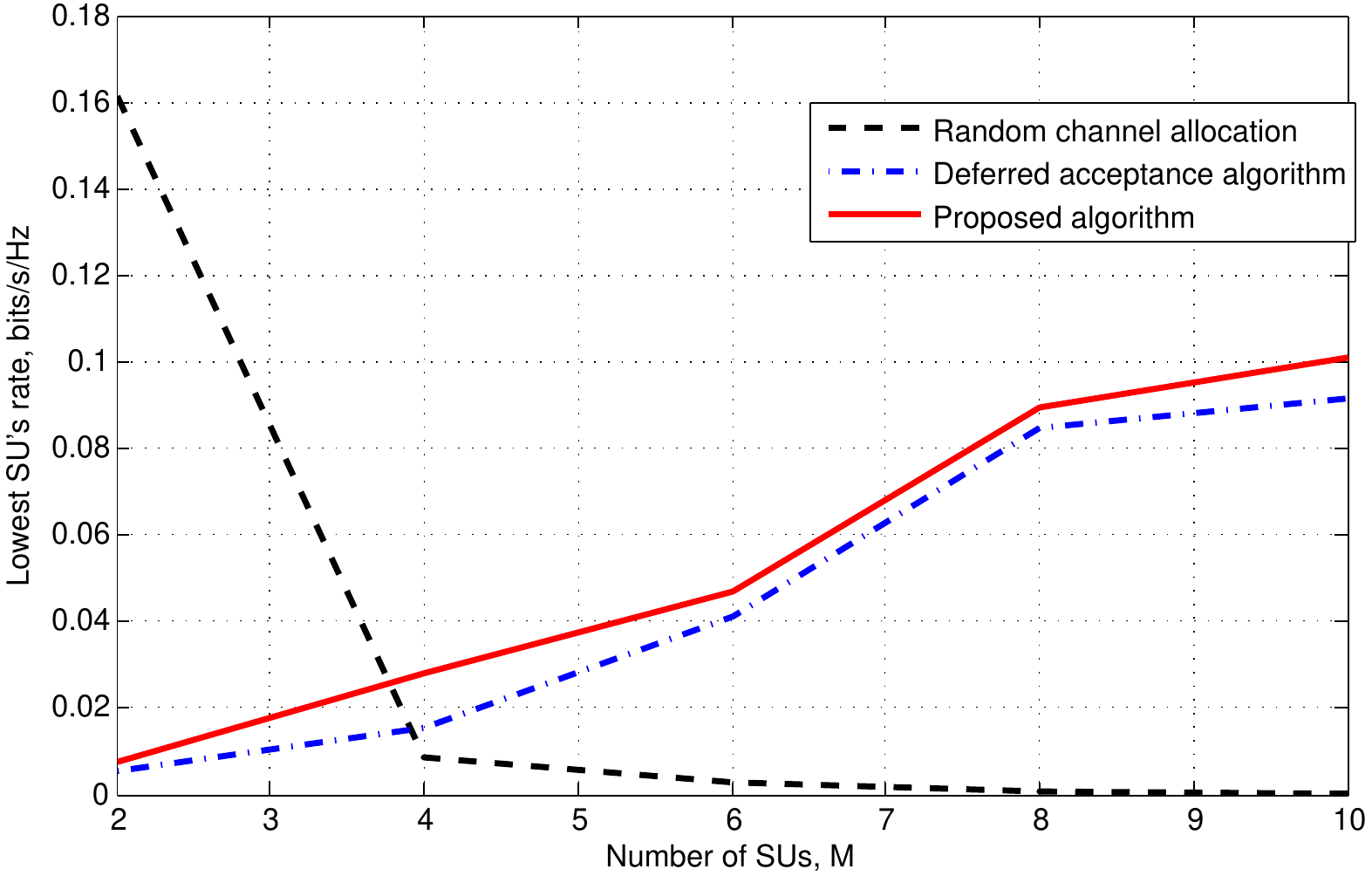}
\caption{\small The lowest rate (worst performance) of an SU with respect to the number of SUs in the network, $M$, for $N$ = $3$.}
\label{fig:SUsworstcase}
\end{figure}
\begin{figure}[t]
\centering
\includegraphics[width=0.65\textwidth,height=0.45\textwidth]{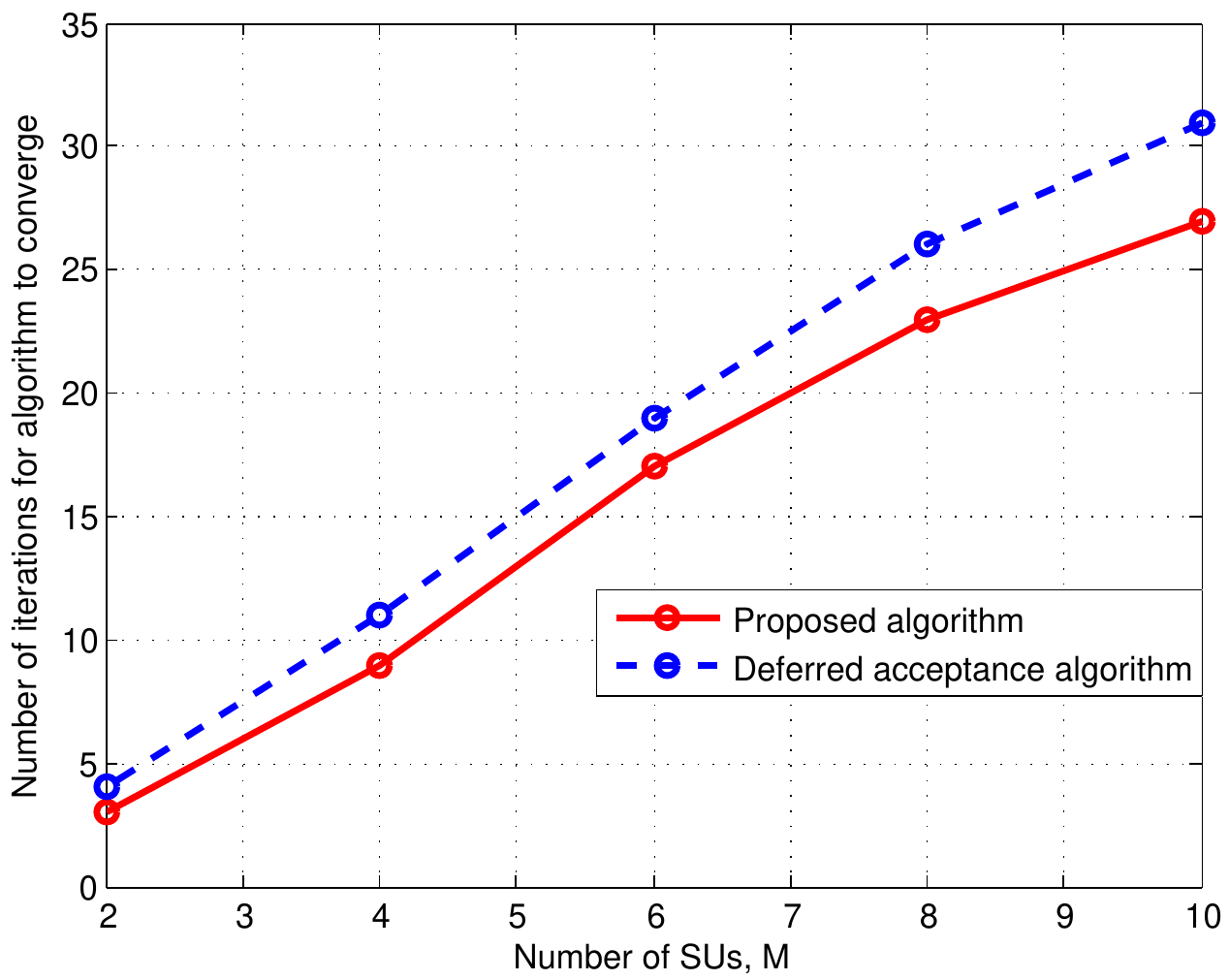}
\caption{\small The number of iterations required for an algorithm to converge with respect to the number of SUs, $M$, when $N = 4$.}
\label{fig:iter-SU}
\end{figure}
For numerical simulations, we consider a CRN having $M = 10$ SUs and $N = 4$ PUs that are uniformly deployed in a square area of $100$ m $\times$ $100$ m. The transmit power of each SU is $13$ dBm and of each PU is $17$ dBm, and the noise level is given such that $\sigma^2$ = $-90$ dBm. 
The path loss exponent is set to $3$. Without loss of generality, we assume $\boldsymbol{\pi}_1 = \cdots = \boldsymbol{\pi}_M$ = $[.1, .2, .3, .4]$. 
Also, note that the results presented below are averaged over a $100000$ simulation runs.
For illustration purposes, we assume that the PUs are inactive and $u_{n,m}(v_{m,n})= 1 -  e^{-v_{m,n}}$. 

In this section, we present results on the performance of our proposed algorithm when compared to the deferred acceptance algorithm \cite{Gale-shap} and a random channel allocation approach. For the deferred acceptance algorithm (explained in our context), SUs include all PUs in their preference lists (in contrast to ours where SUs discard PUs which may lead to negative utilities). For the random channel allocation approach where each SU randomly selects its channel to transmit on, we consider the transmit power of SUs to be twice as much as that considered in our proposed and the deferred acceptance algorithms (assuming that sensing activity in both algorithms consumes the same amount of power as that needed for transmission). 
 
Fig. \ref{fig:throughput-SUs} shows the sum of SUs' rates as the number of SUs $M$ and PUs $N$ vary. In the random channel allocation approach, it is highly likely for a group of SUs to randomly select the same channel to transmit on, especially when $M > N$. As a result, SUs may interfere with each other. Hence, the sum of rates decreases compared to the ones obtained by the deferred acceptance algorithm and our proposed algorithm where each frequency band is allocated to a unique SU. In Fig. \ref{fig:throughput-SUs}, we can see that the proposed algorithm, in terms of the sum of SUs' rates, clearly outperforms both the deferred acceptance algorithm and random channel allocation approach when $M > N$. 
This is due to the following factors: i) higher number of collisions among SUs' transmissions as $M$ increases that is expected to reduce the sum of rates that SUs achieve by following a random channel allocation, and ii) higher number of proposals (due to larger preference lists) to take into consideration when evaluating the value that the sum of SUs' rates converges to when adopting the deferred acceptance algorithm. In the latter case, the sum of SUs' rates is expected to be higher than that of the deferred acceptance algorithm since SUs, according to the proposed algorithm, discard all proposals where $v_{m,n} < 0$. Fig. \ref{fig:throughput-SUs} also shows that the proposed algorithm attains up to $20 \%$ and $60 \%$ improvement in the sum of SUs' rates relative to the deferred acceptance algorithm (for $M = 2$ and $N = 3$) and random channel allocation approach (for $M = 10$ and $N = 4$), respectively.     
  
In Fig. \ref{fig:SUsworstcase}, we show the minimum rate that is achieved by the worst performing SU. 
Based on the proposed algorithm, it is clear that the worst performing SU achieves a minimum rate that increases as $M$ increases. This minimum rate is also higher than what an SU would achieve based on the deferred acceptance algorithm and random channel allocation when $M > N$. The non-decreasing behavior of the proposed and the deferred acceptance algorithms is expected. This is due to the fact that both algorithms try to allocate $N$ bands to only $N$ out of $M$ SUs upon convergence of both algorithms. So, when $M$ decreases, it is expected to either have the same SU achieve the lowest rate or a new SU (with an even worse performance) become part of the stable match. On the other hand, the behavior of the curve corresponding to the worst performing SU following a random channel allocation is anticipated due to large interference power that is created as $M$ increases beyond $N$ which results in a significant decrease in the rates achieved by interfering SUs. Fig. \ref{fig:SUsworstcase} shows that the proposed algorithm achieves up to $25 \%$ and more than $100 \%$ improvement in the worst SU's rate relative to the deferred acceptance algorithm and random channel allocation, respectively.

In Fig. \ref{fig:iter-SU}, we show the number of iterations required for the proposed and deferred acceptance algorithms to converge as $M$ increases for $N = 4$. It is evident from this figure that our proposed algorithm outperforms the deferred acceptance algorithm. This is because the latter algorithm requires more iterations to converge. In other words, actual stabilization of the sum of SUs' rates following the deferred acceptance algorithm takes more time than that of the proposed one because of the larger preference lists that SUs have based on the deferred acceptance algorithm. For example, when $M = 8$, the required number of iterations for the deferred acceptance algorithm to converge is $26$ while our proposed algorithm only needs $23$. 

\section{Conclusion}
\label{sec:Conclusion}

In this paper, we have introduced a novel and distributed model for the allocation of PU-owned frequency bands to the SUs in a CRN. In the proposed model, the SUs sense the licensed spectrum looking for white spaces. We have modeled the problem as a \emph{matching game} between SUs and PUs. We have proposed a novel and distributed algorithm to obtain a \emph{stable match}. We have compared the proposed algorithm with the deferred acceptance algorithm and a random channel allocation approach. Simulation results have shown that the proposed algorithm can improve the sum of SUs' rates of up to $20 \%$ and $60 \%$ relative to the deferred acceptance algorithm and random channel allocation approach, respectively.

An interesting extension, which we shall study later, is to consider a fair and stable association of multiple frequency bands to SUs while accounting for channel imperfections.


\bibliographystyle{IEEEtran}
\bibliography{IEEEabrv,El-Bardan_matching}

\end{document}